\documentclass[useAMS,usenatbib,usegraphicx]{mn2e}

\title[Frame selection techniques]{An investigation of lucky imaging techniques}
\author[A. Smith et al.]{Andrew Smith,$^1$\thanks{E-mail: asmith@science.mq.edu.au (AS)} 
Jeremy Bailey$^2$, J.H. Hough$^3$, Steven Lee$^4$\\
$^{1}$Department of Physics and Electronic Engineering, Macquarie University, NSW 2109, Australia\\
$^{2}$School of Physics, University of New South Wales, NSW 2052, Australia\\
$^{3}$Centre for Astrophysics Research, University of Hertfordshire, College lane,
Hatfield AL10 9AB, United Kingdom\\
$^{4}$Anglo-Australian Telescope, Coonabarabran, NSW 2357, Australia 
}
\begin{document}

\date{Accepted . Received ; in original form }

\pagerange{\pageref{firstpage}--\pageref{lastpage}} \pubyear{2009}

\maketitle

\label{firstpage}

\begin{abstract} We present an empirical analysis of the effectiveness of frame selection
(also known as Lucky Imaging) techniques for high resolution imaging. A high-speed image
recording system has been used to observe a number of bright stars. The observations were
made over a wide range of values of $D/r_0$ and exposure time. The improvement in Strehl
ratio of the stellar images due to aligning frames and selecting the best frames was
evaluated as a function of these parameters. We find that improvement in Strehl ratio by
factors of 4 to 6 can be achieved over a range of D/r$_0$ from 3 to 12, with a slight peak
at $D/r_0 \sim 7$. The best Strehl improvement is achieved with exposure times of 10 ms or
less but significant improvement is still obtained at exposure times as long as 640 ms. Our
results are consistent with previous investigations but cover a much wider range of parameter
space. We show that Strehl ratios of $>$0.7 can be achieved in appropiate conditions whereas
previous studies have generally shown maximum Strehl ratios of $\sim$0.3.  The results are in
reasonable agreement with the simulations of \citet*{baldwin08}. \end{abstract}


\section{Introduction}

The frame selection technique for high resolution imaging involves the recording of a time
series of short exposure images and the selection of the sharpest images out of the series
for alignment and combining into a final image. \citet{fried78} determined that the
probability of obtaining a lucky sharp image (defined as one with wavefront variance less
than $1rad^{2}$) with a telescope of aperture $D$ in seeing described by a Fried parameter
$r_0$ \citep{fried67} is given by:

\begin{equation}
P = 5.6 exp[-0.1557 (D/r_0)^2]
\end{equation}

This suggests that there will be more good quality images available at low $D/r_0$. The probability of
such an image is 1 in 9 for $D/r_0 = 5$ or 1 in 50 for $D/r_0 = 6$. For higher $D/r_0$ the probability
of a sharp image rapidly decreases, being 1 in 3800 for $D/r_0 = 8$. Since the image quality
gain will increase with $D/r_0$ this suggests the frame selection techique will work best
for $D/r_0 \sim 6-7$, this being the largest $D/r_0$ at which there is a good chance
of finding several high quality images in a typical image sequence of a few thousand frames.

There have been a number of practical demonstrations of this technique variously described as
frame selection \citep{roggemann96}, lucky imaging \citep{law06}, or selective image
reconstruction \citep{dantowitz00}. \citet{baldwin01} demonstrated the ability to obtain
diffraction limited star images at 800nm wavelength with a 2.5m telescope. 

The technique has been used to image the hemisphere of Mercury that was missed by Mariner
10 \citep{dantowitz00,cecil07,ksanfomality07} and
is now widely used by amateur astronomers for planetary imaging.

Interest in the technique is rapidly increasing, in part due to the availability of electron
multiplying CCD (EMCCD) technology, which allows rapid readout of CCDs with negligible read
noise \citep{mackay01}, as well as computers with fast processors and large storage
capacity. A number of such systems have recently been demonstrated, for example LuckyCam
\citep{law06}, AstraLux \citep{hormuth08} and FastCam \citep{oscoz08}.

However, previous studies have generally been aimed at obtaining the best possible image
resolution and have therefore explored a restricted range of parameters. In this paper we
present observations that explore a wide range of parameter space. We have explored
empirically the effects of telescope aperture $D$, wavelength $\lambda$, frame exposure
time $t$ and frame selection rate $FSR$ on the resulting image quality. Unlike most
previous studies which have aimed at exploiting excellent seeing conditions, our
observations were obtained in a range of seeing conditions from good to poor. The results
provide information that can help to optimize the design of future instruments.

\section{Observations}

The observations were obtained using \textit{MUSIC} Mk I (Macquarie University Selective
Imaging Camera). The study was carried out as a preliminary stage in the design of a more
advanced lucky imaging system that will use an electron multiplying CCD camera. 
The imager used for \textit{MUSIC} was a Watec 100N monochrome video
camera. This camera was chosen because it had adjustable exposure times and adequate
sensitivity to observe bright stars. The observations used standard BVRI filters. The camera
was placed either directly at the telescope focus, or when necessary used with a 2.5 times 
focal extender. In all cases the image scale was chosen to ensure that the pixels provided
good sampling of diffraction limited star images. 

The camera produces video output with an effective pixel size of 8.6 by 8.3 $\mu$m and
a format of 768 by 576 pixels. The camera generates video data with interlaced scanning (i.e.
each frame consists of consecutive scans of odd rows and even rows stitched together). The
video data was recorded using a Data Translation DT3155 PCI frame grabber mounted in a PC
system using a 3GHz Pentium 4 processor. The PC was configured with 1 TB of disk space (two
400 Gb and one 200 GB drives) to record the large data files. The operating system was
Fedora Core Linux. A data acquisition software system was developed that enabled data from
the video camera to be recorded continuously as three dimensional FITS files, while being
displayed in real time. The software made use of C++ classes developed at the
Anglo-Australian Observatory for the IRIS2 project and AAO2 detector controllers
\citep{shortridge04}. The image display was based on the ESO Real Time Display (RTD) system
\citep{herlin96}. The system was capable of recording full frame video data to disk for
extended periods. Typical observations consisted of sequences of 3000 to 10000 video frames.

\textit{MUSIC} was used on three telescopes. The observing dates, locations and instruments
used are summarised in table~\ref{obssum}. The bulk of the observations were obtained on
the 1m ANU telescope at Siding Spring Observatory. A small number of observations were
obtained on the 3.9m Anglo-Australian Telescope (AAT, also at Siding Spring) and on the 0.4m
telescope of the Macquarie University Observatory in Sydney.

One aim of the observations was to explore the effects of changes in $D/r_0$, which may be
regarded as the seeeing-normalised aperture. While natural variation in seeing provides
changes in $r_0$, we also varied $D$ by placing masks of different size over the telescope
aperture. With the 1m telescope we used mask sizes of 75cm, 30cm, 20cm and 10cm, with the
smaller masks being placed off-centre to avoid the central obstruction of the secondary
mirror. With the AAT off-centre masks corresponding to apertures from 40cm to 1.0m were used
located at the top of the ``chimney'' above the primary mirror central hole. A 2.5m aperture
was also achieved by closing down the primary mirror covers.

The frame exposure times used ranged from $1-20ms$ on the camera. Additional runs with
exposure times from $40-640ms$ were simulated by combining groups of consecutive frames from
the $20ms$ runs. 

Observations were made of a number of bright stars, as well as of some clusters and
binaries of various angular separations to test the effects of selective imaging
on image sharpness across the field. In each case observations were recorded with a range of
different exposure times, and through different filters.

\begin{table} 
\caption{Summary of \textit{MUSIC} Observing Runs} 
\centering \begin{small} 
\begin{tabular}{l l l} \hline
Date	&	Location	&	Instrument\\ \hline 
2005 Mar 8	&  Macquarie University, 	&	Meade 40cm 	\\ &Sydney NSW	&	\\[1ex]
2005 Mar 11-23	&	Siding Spring, NSW	&	ANU 1m	\\[1ex]
2005 Jun 25	&	Siding Spring, NSW	&	AAT 3.9m \\[1ex] 
2005 Nov 9-14	&	Siding Spring, NSW &	ANU 1m	\\[1ex] 
\hline 
\end{tabular} 
\label{obssum} 
\end{small} 
\end{table}

Frames in each FITS cube were calibrated by bias subtraction and flat fielding
(using exposures of the daylight sky). With our camera, the odd and even rows in
each interlaced scan represent two consecitive exposures. Therefore each frame was de-interlaced
by separating the odd and even rows into separate frames and filling the gaps
by interpolation. This produced calibrated FITS cubes with twice as many frames
as the raw ones. Also, in order to maintain precision the new data cubes were
saved in floating point format.

We used a simple peak pixel algorithm \citep{aspin97} for aligning and selecting frames.
This relies on the fact that noise in the image is minimal, which is generally the case for
the bright stars oberved. Different techniques would be needed for fainter guide stars. We
therefore use the value of the highest pixel in each frame as a measure of image quality,
since sharper stellar images will have more flux in the central peak. Post-processing was
done by scanning each cube to find the brightest pixel in each frame, and then the frames
were ranked in order of peak pixel value. The desired fraction of best frames, defined by
the frame selection rate ($FSR$), were aligned so that the position of the peak pixel in
each frame coincided with that in the best frame and then average-combined
(``shift-and-added''). Frame selection rates of 100\%, 10\% and 1\% were used. These were
compared with simulated long exposures made by combining all frames with no alignment
(``stacking''). 

We used the Strehl ratio as a measure of the quality of the resulting star images. Another possible measure is the full width at half maximum (FWHM) of the image. However, this is not, in practice, a good measure of image quality. Images resulting from frame selection generally have a core-halo structure; i.e. a diffraction limited core surrounded by an extended halo the size of the original seeing disk. The core can give rise to a small FWHM even though most of the energy is in the halo. Strehl ratio is a much more demanding measure of image quality, since a high Strehl ratio can only be achieved if most of the energy is in the diffraction limited core. We measured the Strehl ratios using the  ``strehl'' command of the ESO \textit{Eclipse} software package ~\citep{eclipse}. This works by comparing the observed images with a simulated diffraction-limited point spread function, which takes into account, where appropriate, the central obstruction of the aperture in a Cassegrain telescope.

\begin{figure*}
\includegraphics[width=1.0\textwidth]{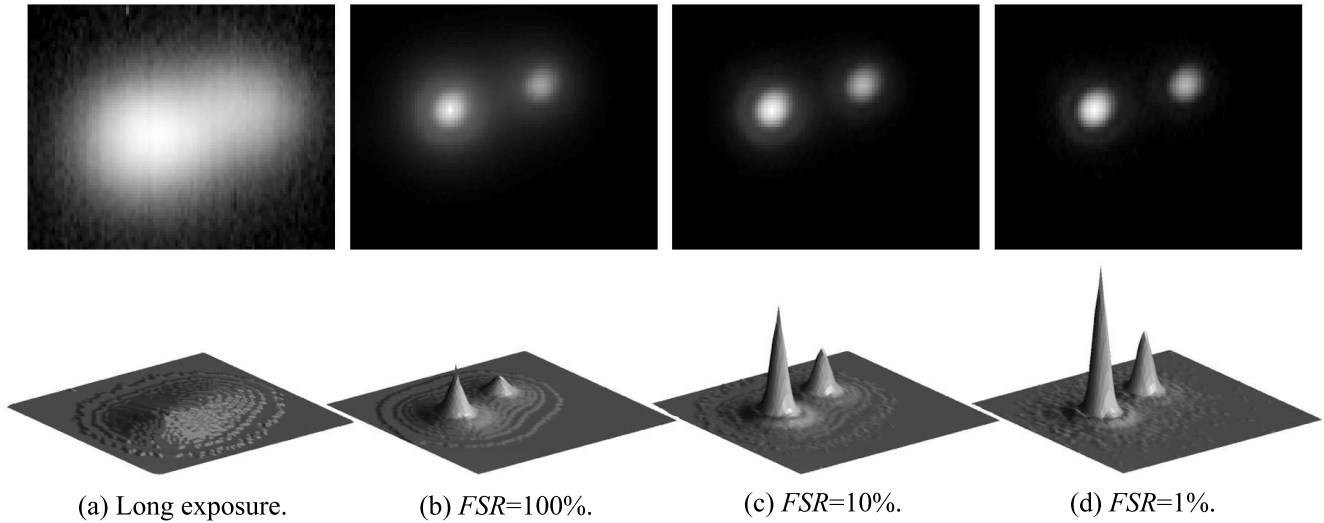}
\caption{Images of binary star $\tau$Ophiuchii (1.7" separation) produced by selective imaging. Images in the top row are brightness-normalised and logarithmic pixel scale to highlight the faint halo. The 3-dimensional plots in the bottom row are linear scale, and show increasing peaks with decreasing $FSR$. Images were taken on the AAT, $D=40cm$ aperture mask, $t=120ms$, $\lambda=0.7\mu$m filter.}
\label{starpics}
\end{figure*}

\section{Discussion}
\subsection{Frame Selection Rate $FSR$}

The stellar images produced by frame selection shown in figure~\ref{starpics}
are among our best results. Note that the images in the top row are displayed
brightness-normalised and logarithmic pixel scale to highlight faint features.
In figure~\ref{starpics}(a) the frames were stacked, simulating a long exposure
image. The next image has all frames shift-and-added ($FSR=100\%$), which is
equivalent to tilt correction. The binary is well resolved with a separation of
1.7 arcseconds that matches published values. Both stars have a clear central
peak with a significant halo. The third and fourth images have the best 10\%
and 1\% of frames aligned and combined. This removed those frames with the
greatest speckling caused by high order wavefront distortions. The result is
greater flux in the central peak and a fainter halo. In the 1\% $FSR$ image
there is a dark ring at a radius of 8 pixels from each peak, matching the
Rayleigh criterion for a diffraction limited image. The 3-dimensional plots
show the diminishing halo and increasing brightness of the peaks with
decreasing $FSR$, which is not shown by the brightness-normalised images. The
Strehl ratios for both peaks in each frame-selected image listed in
table~\ref{starstrehls} verify the quality improvement, with the $FSR=1\%$
image having a Strehl raio of almost 0.8.

The Strehl ratios listed in table~\ref{starstrehls} illustrate an important effect of
shift-and-add processing. In the long exposure image the two peaks are unresolved so no Strehl ratio measurement is possible. In each of the frame selected images the Strehl ratio of the secondary peak is higher than that in the long exposure, but in all cases is less than the corresponding primary peak. This is because even frames with high overall sharpness may still  contain more than one isoplanatic patch. When frames are co-aligned on the brightest pixel the Strehl ratio of the primary peak is artificially enhanced. Other peaks may have a different tilt and speckle pattern. Shift-and-adding may still improve the sharpness of these features, but to a lesser extent than the primary peak, as seen in figure~\ref{starpics} and table~\ref{starstrehls}. Our data shows a general trend to diminishing Strehl ratio and less Strehl improvement with increasing angular distance from the alignment location. This trend seems to be largely independent of $FSR$ and $D/r_0$. This is in qualitative agreement with the simulations by~\citet*{baldwin08}, though a direct quantitative comparison is not possible. However in all of our observations the secondary peaks showed improvement with frame selection, even over angular separations as large as 100 arcseconds.

\begin{table} 
\caption{Strehl ratios of peaks in images in figure~\ref{starpics}} 
\centering \begin{small} 
\begin{tabular}[h]{l c c c c} \hline
	&	\multicolumn{3}{c}{Frame Selection Rate}	\\	
Peak	&	100\%	&	10\%	&	1\%	\\	\hline
Primary	&	0.517	&	0.654	&	0.799	\\	[1ex]
Secondary	&	0.362	&	0.547	&	0.674	\\ [1ex]
Secondary/Primary & 69.9\% & 83.7\% & 84.4\%	\\
\hline 
\end{tabular} 
\label{starstrehls} 
\end{small} 
\end{table}

\subsection{Seeing-normalised aperture $D/r_0$}
\begin{figure}
\includegraphics[height=86mm,angle=270,viewport=60 0 500 700,clip]{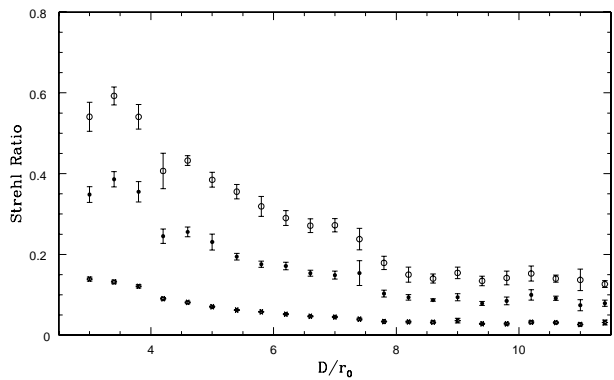}
\caption{Strehl ratio versus $D/r_0$ of stellar images binned in $D/r_0$. Open circles are the results of selecting
and aligning the best 1\% of frames. Filled circles are the results of aligning all images with no frame
selection (``shift-and-add" processing). Stars are the results of summing frames with no shifts
(i.e. giving the equivalent long exposure image).}
\label{strehlDr0}
\end{figure}

We determined $r_0$ from the FWHM of the long exposure image using the standard relationship $FWHM = 0.98 \lambda/r_0$. Over all of our observations $r_0$ ranged from 1 to 12cm, and $D/r_0$ from 2.8 to 30.3. Figure~\ref{strehlDr0} plots bin-averaged Strehl ratios of stellar image versus $D/r_0$ between 3.0 and 12.0. For $D/r_0>12.0$ the trends continue more or less flat. For $D/r_0<12.0$ the obvious trend is for higher Strehl ratios in better seeing (large $r_0$) and/or with smaller aperture (small $D$). With low $D/r_0$, not
only are there more good frames to choose from (one could use a quality threshold instead of frame selection rate) but the average frame quality is better, as indicated by higher Strehl ratios of the long exposure images in this region. This makes frame selection most suitable for use with small to medium sized telescopes at visible wavelengths. This does, however, limit the magnitudes of usable target objects or guide stars.

\begin{figure}
\includegraphics[height=86mm,angle=270,viewport=60 0 500 700,clip]{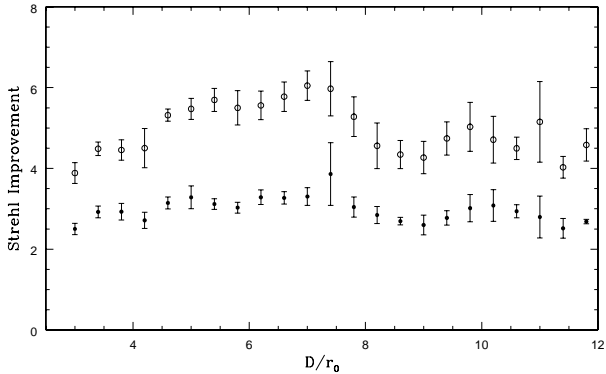}
\caption{As for figure~\ref{strehlDr0}, but plotting the improvement in strehl ratio compared with the long exposure
image for each bin of $D/r_0$. Open circles are the results of selecting
and aligning the best 1\% of frames. Filled circles are the results of aligning all images with no frame
selection (``shift-and-add" processing). }
\label{strehlimprov}
\end{figure}

Figure~\ref{strehlimprov} shows the quality improvement versus $D/r_0$. The improvement factor was measured by the gain in the Strehl ratio; that is, the Strehl ratio of a frame selected image divided by that of the long exposure image derived from the same image cube. Figure~\ref{strehlimprov} compares the improvement made by pure shift-and-add ($100\% FSR$) with that from $1\% FSR$. The $1\% FSR$ points are consistently higher than those for $100\% FSR$, confirming the advantage of being more selective. The $1\% FSR$ images show an improvement factor greater than 5 for $D/r_0$ between 4.5 and 7.8, with a small peak at $D/r_0\sim7$. This represents the best compromise between the diffraction-limited resolution, which improves with increasing $D$, and seeing-limited resolution that improves with increasing $r_0$. Nevertheless the peak is not very pronounced, and substantial Strehl ratio improvement is obtained at all values of $D/r_0$.

\begin{figure}
\includegraphics[width=90mm]{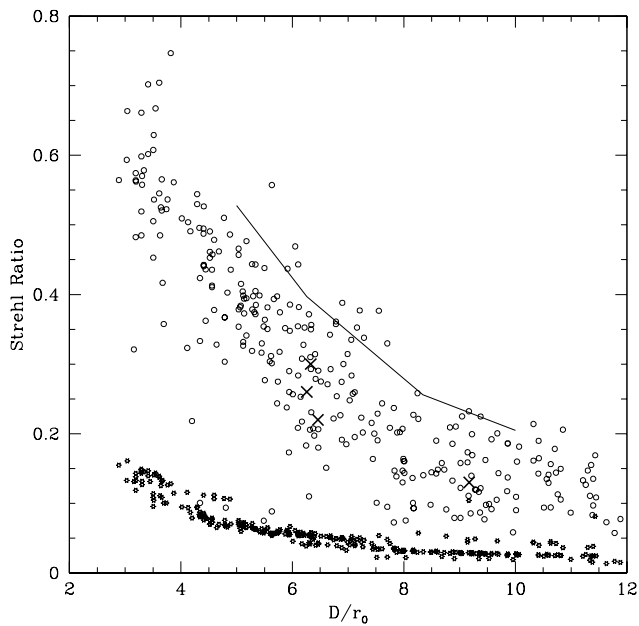}
\caption{Strehl ratio versus $D/r_0$ for individual observations with exposure time of 4 ms or less. Open
circles are the result of selecting and aligning the best 1\% of frames. Stars are the reult
of summing all frames without alignment or selection giving the equivalent long exposure image.
The line is the prediction from the simulation of \citet{baldwin08}. Crosses are previous 
lucky imaging results as described in the text.}
\label{strehlscat}
\end{figure}

Data from our individual observations is displayed in figure~\ref{strehlscat} compared with results from other experiments. In this plot our data are limited to $ t\leq4ms$ and $D/r_0\leq 12$, $1\%$ and $100\% FSR$. The crosses are previous lucky imaging results \citep{baldwin01, law06, hormuth08} and match our data well, even though our data was generally obtained in poorer seeing but with smaller apertures. However, these previous studies generally targeted the optimum case of $D/r_0 \sim 7$, and acheved maximum Strehl ratios of $\sim 0.3$. Our data show that smaller $D/r_0$ values can be used to achieve  higher strehl ratios of $>0.6$ and in a few cases as high as 0.8. The high Strehl ratios for the images in figure~\ref{starpics} were achieved with $D/r_0=3.84$.

The line in figure~\ref{strehlscat} is the simulation from \citet{baldwin08}. It can be seen that the simulation line lies at the upper boundary of the scatter of points. Typically both our observations and previous results lie below the simulation. This is probably due to aberrations in the optics of the telescopes employed. The simulated case assumes a diffraction limited telescope. In this case frame selection will select those images in which the wavefront distortion due to turbulence is minimal. When using a real telescope with aberrations it is necessary instead to select those frames in which the turbulence induced wavefront distortions cancel out those due to telescope aberrations. The probability of a "lucky image" in this case is lower than that for a perfect difraction limited telescope \citep{beckers96} and hence the image quality gain from frame selection is reduced. 

\subsection{Colour band $\lambda$}
Because $ r_{0}\propto\lambda^{6/5} $ \citep{fried66} it was expected that the better Strehl ratios would be achieved at longer wavelengths due to the larger $r_0$ patch. Table~\ref{colours} shows the ranges of $r_0$, $D/r_0$ and long exposure (stacked) Strehl ratios measured from our obseravtions in the V-band (0.55 $\mu$m) and I-band (0.80 $\mu$m). Because of the lower range of $D/r_0$ in the I-band images their average frame quality was better than in other bands, giving higher average Strehl ratios in both the long exposures and the frame selected images. A smaller number of observations obtained in the B and R bands were consistent with this trend.

\begin{table} 
\caption{Ranges of $r_0$, $D/r_0$ and long exposure Strehl ratios measured in the V and I bands.} 
\centering \begin{small} 
\begin{tabular}{l c c c c c c} \hline
&	\multicolumn{2}{c}{ $r_0 (cm)$} &	\multicolumn{2}{c}{ $D/r_0$} &\multicolumn{2}{c}{ Long exp. } \\ 
& & & & &\multicolumn{2}{c}{Strehl ratios} \\
\hline
Band	&	V	&	I	&	V	&	I	&	V	&	I \\ \hline
max	&	7.3	&	12.2	&	25.90	&	22.72	&	0.107	&	0.161	\\
min	&	1.2	&	2.1	&	4.64	&	2.89	&	0.008	&	0.008	\\[1ex]
\hline
\end{tabular} 
\label{colours} 
\end{small} 
\end{table}


\subsection{Frame exposure time $ t $} To effectively freeze the turbulence the frame exposure time must be less than $t_{0} \sim r_0/v$, where $v$ is the bulk wind velocity in the region responsible for the turbulence. However a small $t$ limits the available targets to the brightest objects, and reduces the $ SNR $ of each frame. Thus it is crucial in frame selection to optimise $t$.

\begin{figure}
\includegraphics[width=94mm]{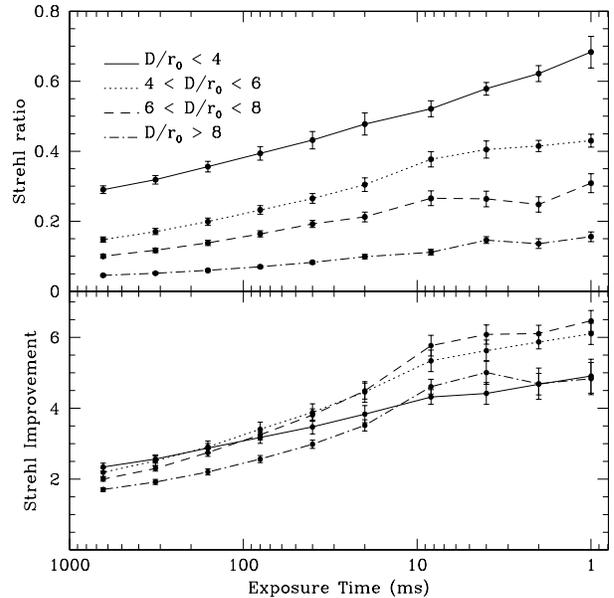}
\caption{Strehl ratio (upper panel) and Strehl improvement factor (lower panel) for data processed
by selection and alignment of the best 1\% of frames as a function of exposure time. The data are
binned into four different ranges of $D/r_0$}
\label{strehlt}
\end{figure}

The top panel of figure~\ref{strehlt} shows Strehl ratios against exposure times for $1\%
FSR$, with the data binned into four ranges of $D/r_0$. The bottom panel plots the improvement in Strehl ratio over the stacked images. Even at the longest times tested ($640ms$) the Strehl ratios improved by a factor of 2. Shorter times gave greater improvement, but the curves appears to flatten below 8 to $10 ms$, especially for larger $D/r_0$. Hence we find that, although selective imaging gives improved image quality for any reasonably short exposure time, if the target is sufficiently bright $t$ should be limited to $10ms$. For our typical $r_0$ of
about 5 cm this corresponds to a wind speed of 5 $ms^{-1}$. The ground wind speeds measured during the runs at Siding Spring were typically in the range 3-8 $ms^{-1}$ and therefore consistent with a significant amount of the seeing being generated near the ground. In better seeing or at longer wavelengths longer frame exposure times should be acceptable.

\section{Conclusions}

Our analysis confirms previous results in showing that substantial improvements in image
Strehl ratio can be achieved by selecting and aligning the sharpest frames in a time series
of short exposure images. By reducing the telescope aperture to be a few multiples of $r_0$, thereby
minimising $D/r_0$, and by selecting only 1\% of the best quality frames Strehl ratios as high
as 0.6 to 0.8 were obtained. The improvement was greatest when imaging at longer wavelengths due to the larger values
of $r_0$. The optimum gain in Strehl ratio over long exposures was found to be about a factor
of 6 at $D/r_0\sim
7$. However, the Strehl gain is rather insensitive to D/$r_0$ with improvements of a
factor of 4 or more being obtained over the full range of D/$r_0$ from 3 to 12. The
improvements obtained by aligning frames without any selection (shift-and-add processing)
are smaller, ranging from 2 to 3.

Frame selection has been found to improve image sharpness over a wide range of frame
exposure times. Even with exposure times as long as $640 ms$ the Strehl ratio was improved by a factor of 2. However the best gains in Strehl ratio (from 4 to 6) for the
Siding Spring site are obtained with $t<10ms$. The optimum exposure time may be longer for
sites with better seeing or at longer wavelengths. 

Our results are consistent with previous lucky imaging studies, but explore a wider range of
parameter space and show the potential for achieving larger Strehl ratios than previous
results, which have generally been limited to Strehl ratios $<$0.3. The variation of Strehl
ratio with D/$r_0$ we observe shows a similar form to, but lies below the simulations of
\citet{baldwin08}. This is probably due to telescope aberrations leading to a lower
probability of achieving a lucky sharp image.

\section*{Acknowledgments}

We thank the staff of the Siding Spring Observatory and the Anglo-Australian Telescope for their suport of our observations. We also thank the Director of the Anglo-Australian Telescope, Matthew Colless, for provision of time on the AAT, and the RSAA Time Allocation Committee for assignments of time on the ANU 1m telescope. The helpful suggestions from Alan Vaughan and Mark Wardle of Macquarie University are also appreciated.

\label{lastpage}


\begin{thebibliography}{}

\bibitem[\protect\citeauthoryear{Aspin et al.}{1997}]{aspin97} Aspin, C., Puxley, P.J.,
Hawarden, T.G., Paterson, M.J., Pickup, D.A., 1997, MN, 284, 257.
\bibitem[\protect\citeauthoryear{Baldwin et al.}{2001}]{baldwin01} Baldwin, J.E., Tubbs,
R.N., Cox, G.C., Mackay, C.D., Wilson, R.W., Andersen, M.I., 2001, A\&A, 368, L1
\bibitem[\protect\citeauthoryear{Baldwin, Warner \& Mackay}{Baldwin et al.}{2008}]{baldwin08} Baldwin, J.E.,
Warner, P.J., Mackay, C.D., 2008, A\&A, 480, 589
\bibitem[\protect\citeauthoryear{Beckers \& Rimmele}{1996}]{beckers96} Beckers, J.M. and
Rimmele, T.R., 1998, BAAS, 28, 1325
\bibitem[\protect\citeauthoryear{Cecil \& Rashkeev}{2007}]{cecil07} Cecil, G. \&
Rashkeev, D., 2007, AJ, 134, 1468
\bibitem[\protect\citeauthoryear{Dantowitz, Teare \& Kozubal}{2000}]{dantowitz00}
Dantowitz, R.F., Teare, S.W., Kozubal, M.J., AJ, 119, 2455
\bibitem[\protect\citeauthoryear{Devillard}{2001}]{eclipse} Devillard, N., 2004, in Harnden, Jr., F.~R. and Primini, F.~A. and Payne, H.~E. (eds) Astronomical Data Analysis Software and Systems X, Astronomical Society of the Pacific Conference Series, 238, 525-528
\bibitem[\protect\citeauthoryear{Fried}{1966}]{fried66} Fried, D.L., 1966, J. Opt. Soc. Am., 56, 1372-1379
\bibitem[\protect\citeauthoryear{Fried}{1967}]{fried67} Fried, D.L., 1967, Proc. IEEE, 55, 57  
\bibitem[\protect\citeauthoryear{Fried}{1978}]{fried78} Fried, D.L., 1978, J. Opt. Soc. Am., 68, 1651
\bibitem[\protect\citeauthoryear{Herlin, Brighton \& Biereichel}{1996}]{herlin96} Herlin,
T., Brighton, A., Biereichel, P., 1996, in Jacoby, G.H , Barnes, J. (eds), Astronomical Data
Analysis Software and Systems V, A.S.P. Conf series, 101, 396
\bibitem[\protect\citeauthoryear{Hormuth et al.}{2008}]{hormuth08} Hormuth, F., Brandner,
W., Hippler, S., Henning, Th., 2008, in Schoedel, R., Eckart, A., Pfalzner, S., Ros, E.
(eds) The Universe under the Microscope - Astrophysics at High Angular Resolution, J. Phys.
Conf. Series, 131, 012051
\bibitem[\protect\citeauthoryear{Kern et al.}{2000}]{kerntemp} Kern, B., Laurence, T.~A., Martin, C. and Dimotakis, P.~E., 2000, App. Opt. 39, 4879-4885
\bibitem[\protect\citeauthoryear{Ksanfomality \& Sprague}{2007}]{ksanfomality07}
Ksanfomality, L. \& Sprague, A.L., 2007, Icarus, 188, 271
\bibitem[\protect\citeauthoryear{Law, Mackay \& Baldwin}{Law et al.}{2006}]{law06} Law, N.M.,
Mackay, C.D., Baldwin, J.E., A\&A, 446, 739
\bibitem[\protect\citeauthoryear{Mackay et al.}{2001}]{mackay01} Mackay, C.D., Tubbs, R.N.,
Bell, R., Burt, D.J., Jerram, P., Moody, I., 2001, in Morley, M.B., Canosa, J., Sampat, N.
(eds), Sensors and Camera Systems for Scientific, Industrial and Digital Photography
Applications II, Proc SPIE, 4306, 289
\bibitem[\protect\citeauthoryear{Oscoz et al.}{2008}]{oscoz08} Oscoz, A. et al., 2008, in
McLean, I., Casali, M. (eds) Ground-based and Airborne Instrumentation for Astronomy, SPIE
Proc., 7014, 701447
\bibitem[\protect\citeauthoryear{Roggemann \& Welsh}{1996}]{roggemann96} Roggemann, M.C. \&
Welsh, B., 1996, Imaging through Turbulence (Boca Raton: CRC Press)
\bibitem[\protect\citeauthoryear{Shortridge et al.}{2004}]{shortridge04} Shortridge, K.,
Farrell, T.J., Bailey, J.A., Waller, L.G., 2004, in Lewis, H., Raffi, G. (eds) Advanced
Software, Control and Communications Systems for Astronomy, SPIE Proc., 5496, 463



\end{thebibliography}
\end{document}